# Ingrained: an automated framework for fusing atomic-scale image simulations into experiments

Eric Schwenker[1,2], V.S. Chaitanya Kolluru[1,3], Jinglong Guo[4], Xiaobing Hu[2], Qiucheng Li[2], Mark C. Hersam[2], Vinayak P. Dravid[2], Robert F. Klie[4], Jeffery R. Guest[1], Maria K. Y. Chan[1*]

1. Center for Nanoscale Materials, Argonne National Laboratory, USA
2. Department of Materials Science and Engineering, Northwestern University, USA
3. Department of Materials Science and Engineering, University of Florida, USA
4. Department of Physics, University of Illinois Chicago, USA

corresponding author: Maria Chan (mchan@anl.gov)

## Abstract

To fully leverage the power of image simulation to corroborate and explain patterns and structures in atomic resolution microscopy (e.g., electron and scanning probe), an initial correspondence between the simulation and experimental image must be established at the outset of further high accuracy simulations or calculations. Furthermore, if simulation is to be used in context of highly automated processes or high-throughput optimization, the process of finding this correspondence itself must be automated. In this work, we introduce *ingrained*, an open-source automation framework which solves for this correspondence and fuses atomic resolution image simulations into the experimental images to which they correspond. We describe herein the overall *ingrained* workflow, focusing on its application to interface structure approximations, and the development of an experimentally rationalized forward model for scanning tunneling microscopy simulation.

## Introduction

Materials image simulations are becoming an integral part in the structural analysis of complex materials systems. Having a three-dimensional atomistic structure of the system under study is valuable, both for understanding and for property prediction through first principles simulations. For transmission electron microscopy (TEM) and scanning TEM (STEM), the electron-matter interactions intrinsic to the image formation process are well-codified in numerical "multislice" simulations [1], and the combination of aberration-corrected STEM images with these multislice simulations have been used effectively in a variety of contexts for structural determination with atomic precision [2]–[7]. Image simulations have also proven useful in scanning tunneling microscopy (STM) in order to help solve for surface structure or adsorption geometries [8]–[11]. The success of these "simulation to experimental" comparisons is rooted in their ability to link information about the underlying mechanisms generating the experimental observation to parameters and/or specific structures used in simulation. However, to utilize simulations for this purpose, a mapping between the simulation and the visual or measurable expectation from experiment must be explicitly established (*i.e.,* pixels from one image are mapped to corresponding pixels in another). The process of establishing this "correspondence" is referred to as image registration.

Image registration workflows are often divided into coarse and fine alignment steps [12]. Coarse alignment reduces the search range in the subsequent fine alignment by bringing the two images into rough coincidence. For real-space structural characterization of materials using microscopy, particularly when imaging crystalline solids at high resolution, coarse alignment usually involves reducing the search space to a smaller, more tractable area. Even when the image contains an interface [6], comparisons between simulation and experiment are typically confined to a reduced field of view, acting as feature substitute for the more complex regions of the image.

A slightly more automated approach to coarse alignment involves identifying salient points and their correspondences (so-called *landmarks*) across images. With landmarks in place, a point-set registration algorithm such as the iterative closest point (ICP) [14] ensures that the distance between corresponding landmarks is minimized, and thus roughly aligned. Landmark identification is used in both *pycroscopy* [17], and the 'TurboReg' plugin for ImageJ [18] (image processing platforms for microscopy) as a recommended initial step before full registration is executed. There are ways to automate the selection and pairing of landmarks borrowing from computer vision (SIFT + RANSAC [19], [20]), but this is generally expensive. In addition to landmarks, intensity correlation is another approach to semi-automated coarse alignment, and fortunately, it does not rely on artificial markers placed in the field of view. In the simplest cases, an approach such as phase cross-correlation [21] can automatically remove translation (and rotation [22]) offset in images collected from pre-aligned sensors, or from images collected in rapid succession as part of an image stack (video). These are the initial steps taken by the popular SmartAlign [23] tool, which provides general-purpose image processing for atomic-resolution series data from STEM.

In cases where registration is cast as a two-step procedure, an affine linear transformation is often sufficient for initial coarse alignment. This assumes that major spatial discrepancies between images can be corrected by a combination of rotation, translation, scaling, and shear. For the final "fine alignment" step, a straightforward intensity-based approach proceeds as an iterative optimization of a "similarity measure" which takes into account the explicit pixel values in each image [24], and in some cases, even subpixel shifts [25], [26]. If alignment transformation cannot be expressed with rigid deformations, a non-rigid registration approach can resolve local discrepancies in image content with a set of local deformations. Non-rigid registration has been used for purposes ranging from scan instability corrections in STEM imaging [27], to registration of MRI brain images to help capture brain shift during surgery [28]. The above examples use registration to compensate for spatial discrepancies that exist between two images, with the assumption that the objects in the images are similar enough to be overlayed on top of each other. Here registration can provide a quantitative measure of how the structures differ, or a means to direct comparison. Contrast this with registration needs for simulated materials characterization images that are used to corroborate experimental finding. With this, the goal often involves more than just a solution for a single spatial transformation, but also, the flexibility to modify the structure and or/imaging parameters to the forward model in the loop (*i.e.,* structure and parameter iteration are valuable additions to the overall registration framework).

In most standard contexts, registration of a simulated materials characterization image with an experimental image is considered separate from the forward simulation itself. This is a sufficient approach when enough is known about the parameters of the forward model to produce a reasonably appropriate image, but this is often not trivial. For example. STM measurements are routinely used to probe thin film surface morphology at atomic resolution. Simulations of STM images from density functional theory (DFT) charge densities using the Tersoff and Hamann approximation [29] often accompany these measurements in order to explore various surface geometries in a systematic way, but can be challenging to match to experiment because the overall appearance of the image is greatly influenced by small changes in parameters such as the charge density value to construct the isosurface or the vertical distance of orbitals below the surface that are considered to be accessible by the STM tip. These parameters are difficult to determine quantitatively from experimental conditions and DFT results, and thus a decoupled forward modeling and registration paradigm involving manual trial-and-error is less than ideal for these characterization techniques.

In this paper we introduce *ingrained*, an automated framework for image registration which allows for the fusion of atomic-resolution materials imaging simulations into the experimental images to which they correspond. The framework is modular, allowing for plug-and-play implementation of forward models for image simulations when an experimental complement exists, and provides tools for programmatic construction of periodic bicrystal interfacial structures from materials database queries. In addition to a framework overview, we outline two valuable use cases for image registration with *ingrained*: (1) an experimentally-informed initial bicrystal structure for further interface structure refinement through heuristic search algorithms (e.g. basin hopping or genetic algorithm), or high-accuracy structure refinement with multislice and high angle annular darkfield (HAADF) STEM comparisons; and (2) an experimentally-rationalized forward model for STM image simulation that involves fine-tuning of imaging parameters. The ingrained toolkit has recently been used to determine the structure of rectangular hydrogenated borophene, synthesized for the first time, from STM images [30]. Examples and instructions for access for ingrained are available on GitHub (https://github.com/MaterialEyes/ingrained-lite).

**Methods**

*Framework Overview.* The *ingrained* framework requires an experimental atomic-resolution microscopy image as input. Conventional preprocessing operations such as Wiener filtering are often useful for simple restoration of the original image if it has undergone significant degradation during acquisition. The main components of the *ingrained* framework are depicted in Figure 1. The first step is structure initialization, where initial input parameters are used to assemble a starting structure (a bicrystal in this instance, but other non-interface structures are possible). Next, a forward model produces a simulated image of the starting structure, which is fused with the experimental image after an iterative optimization procedure workflow, outside of potential image preprocessing, involves setting up the parameters and/or constraints used for structure initialization and forward modeling - no landmarks or manual image manipulation are necessary to achieve suitable image registration. The final registration result is a simulated image with a parameterized fit-to-experiment, as well as the structure approximation (*i.e.,* the structure that was used as the basis for the matching simulation). The following sections provide additional input and implementation details.

*Structure Initialization.* Depending on the nature of the imaged structure and its associated imaging modality, the *ingrained* framework offers two methods for structure initialization. In the simplest case, a database query tool can be used to programmatically download chemical structure files from the Materials Project (MP) database [31] based on *chemical formula* and *space group* information provided by the user in a configuration file (JSON format). With this, if the goal is to register a bulk crystalline structure to the experimental image and the user specifies a viewing direction, the structure is also rotated to the prescribed orientation as part of the initialization. The second method of structure initialization involves the automatic construction of bicrystal interfaces. In this mode, the configuration information is used to detail the composition and viewing orientation for two crystal structure files (grains) in the image, as well as constrain certain dimensions of the "over lattice" constructed around the overall composite bicrystal. In both the single structure query and the bicrystal construction, the user specifies orientation by providing the *uvw_projection* direction (*i.e.,* direction from viewpoint to screen), the *uvw_upwards* direction (*i.e.* upwards direction on the screen) and the *tilt angle* which is a misorientation applied after the previous vector constraints have been satisfied. The required construction parameters include: *min/max depth, min/max width, interface width, etc.* Currently, the size of the over lattice (which defines the simulation cell for the bicrystal) is controlled by these restrictions on the real-space dimensions. Image recognition tools aimed at identifying chemistry, scaling, and orientation information from the images directly is an avenue of further research that could enhance the automation in this step.

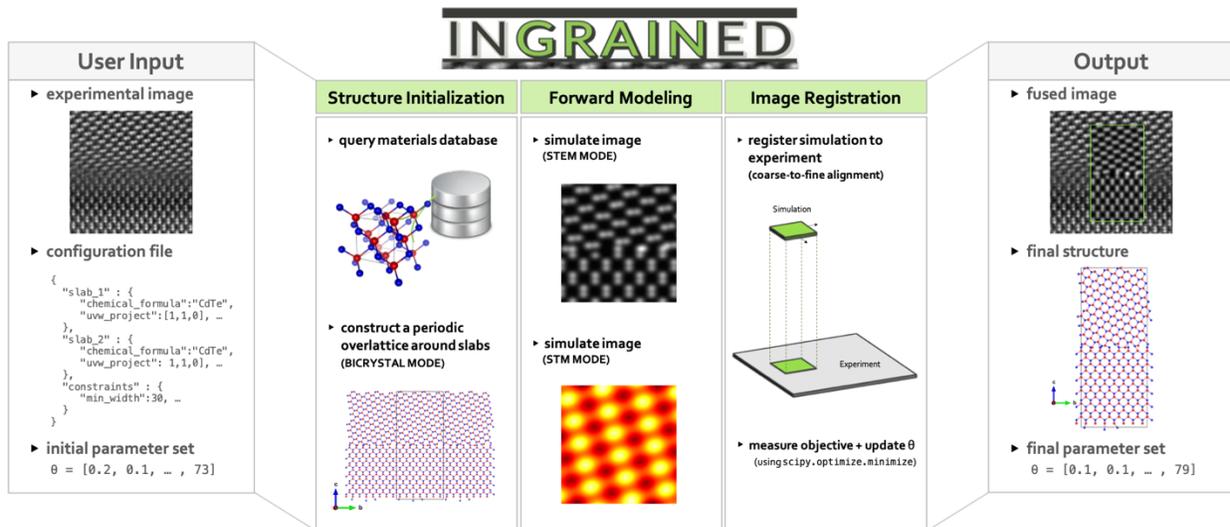

**Figure 1. | Overview of the *ingrained* framework.** This is the setup for the *ingrained* framework when the experimental microscopy image contains an interface. In addition to an atomic resolution image, a user is required to input information about the chemistry and orientation of the grains, as well as an initial guess for the parameter set, which together control both the structure initialization and forward modeling components. The structure initialization module (in bicrystal mode) is used to query a materials database for the structures of the grains, and then constructs a periodic over lattice around the grains which are combined into a bicrystal supercell. A forward modeling option (both STEM and STM are implemented) is selected and an image is simulated. From there, the image registration module iteratively optimizes the parameter set for the forward model, to produce the simulation with the best fit inside the experimental image. The 'image fusion' result is available as output, alongside the final structure simulation which contains a parameterized fit-to-experiment.

Proceeding in this bicrystal mode, the information in the configuration file is used to construct two oriented grains that are then combined into a single bicrystal structure, satisfying periodicity conditions through application of uniform strain (*i.e.,* small discrepancies in individual dimension requirements are removed by strain). This is necessary, particularly for interface structures with low symmetry, as it is otherwise intractable to create atomic structures that are small (for computational efficient in simulations) but remain periodic in at least two dimensions (required for some simulation approaches). The procedure for ensuring periodicity involves estimating repeat length from the grains, and then using the near-coincidence site lattice approach (CSL) with subspace search outlined in Buurma *et al.* [32] to determine the appropriate dimensions for the subsuming "over lattice". The bicrystal can then be used in materials modeling context and for image simulation.

*Forward Modeling.* All the previous configuration parameters are specific to the assembly of the initial structure. Note that this structure initialization feature is not required and can be bypassed in situations where an initial structure and/or partial charge density data (STM simulations) is already provided. The relevancy of this initialization step is entirely dictated by the input structure requirements of the forthcoming forward modeling step. In general, the forward model simulates an image from an atomistic structure and requires a set of simulation parameters. These simulation parameters are kept separate from the structure initialization parameters (where applicable) because they are specific to the forward model being implemented.

Currently, the ingrained toolkit provides forward modeling options for both high angle annular darkfield (HAADF) STEM image simulation, as well as for STM. In the current implementation, HAADF STEM image simulation is performed as a simple convolution of the atomic coordinates with a point spread function for the microscope, using Kirkland's incostem code [33]. The specific parameters that control image formation based on physical principles (i.e., defocus, sample thickness, etc.) are consistent with the parameters discussed in [33]. This convolution approach is convenient because the calculation of an image is performed as a simple multiplication. This provides a tremendous speed advantage over other more quantitatively accurate techniques and in many cases, is capable of capturing many of the same pertinent features [33]. Alternative STEM simulation codes such as Prismatic [34], [35] can also be interfaced with ingrained.

For the STM mode, we developed and implemented a forward model which generates a simulated STM image from the electronic charge densities data within ingrained. Experimentally, the constant current STM images are obtained by moving the tip in parallel lines across the surface while the tip height is adjusted height to maintain a constant current using a feedback loop. Based on the Tersoff and Hammann approximation [29], the surface charge densities near the Fermi level correlate to the tunneling current observed in the experimental STM images. The simulated STM images generated by this forward model are the isosurfaces of charge densities near the Fermi level plotted when observed from the top view. For STM simulation of a surface slab, the energy-selected charge density file from a DFT calculation (in the PARCHG file format in VASP [36][37]) which contains the volumetric data of the partial charge densities in the entire slab structure is the only required input.

The STM forward model requires four variables to simulate an STM image which are the electron density value ($r\_val$) which corresponds to the constant current and a tolerance ($r\_tol$), and the vertical distance below the surface ($z\_below$) and above the surface ($z\_above$) within which the electron densities are assumed to be available to the STM tip. The simulated STM image shows the x-y grid points where the electron densities are within the specified isosurface density range ($r\_val \pm r\_tol$) and the brightness at each grid point corresponds to its height. Several distinct simulated STM images can be generated from one PARCHG file by using different combinations of these input parameters. Ingrained performs constrained optimization of these parameters together with image processing parameters such as shear and strain in both x and y directions, pixel size, image crop size and sigma for gaussian blur to provide a final STM simulation.

*Image Registration*. With *ingrained*, image registration is cast as an iterative optimization problem. It is assumed that scaling discrepancies between the experiment and simulation are minimal, which is reasonable when a raw microscopy image containing scaling metadata is available. The iterative optimization is local in the sense that it is restricted to points that are close to the initial guess, so a multi-start approach is recommended.

After an image is simulated, the first step of image registration is coarse alignment, which implies that both the simulated and experimental images are downsampled and quantized. The alignment step itself is actually an iterative procedure that takes patches of the experimental image that are highly correlated with and the same size as the simulation, and attempts to find one with zero translative offset relative to the simulation. The translative offset is computed using an efficient phase cross-correlation function from scikit-image [38], which finds a position of maximum correlation between the two images in the frequency domain (i.e. maximum correlation yields minimum translative offset). In this step, it is assumed that two atomic resolution images presented at the same scale with no translative offset between them, is a sufficient proxy for geometric consistency at the boundary of the simulation and experimental image. This is based on observation and holds for all cases presented in the results. If an experimental patch is found that satisfies the conditions of coarse alignment, a fine alignment step is applied to the simulation and experiment at native (higher) resolution. The purpose of this step is to search experimental patches in the local vicinity of the matching coordinates from coarse alignment, to find the higher resolution experimental patch with minimum translative offset. After the fine alignment step, the quality of the registration is assessed based on a custom objective function and the entire process is repeated for a new parameter set until the objective satisfies the convergence criteria set by the overall optimizer. Powell's method [39] is the default optimizer for registration, but, in theory, any derivative-free optimization method included as part of SciPy [40] could be used with very minimal revision to the current setup.

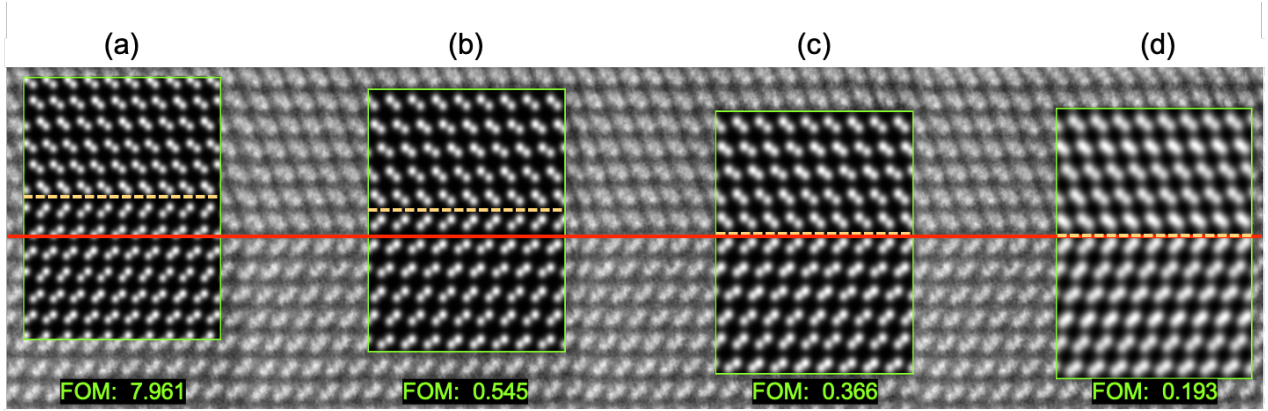

**Figure 2. | Figure-of-merit (FOM) score and fit-to-experiment.** The sequence of images illustrates how the FOM values are indicative of the fit-to-experiment (with the default $\alpha = 0.1$, $\beta=1$ weights). **(a)** For FOM values >1, there is often significant translative offset with respect to the simulation/experiment boundary ($d_{trans}$) since the $d_{sim}$ contribution is bounded on the interval [0,1]. In this case it is due to a scaling discrepancy. **(b)** For FOM values between 0.5 and 1, usually one or both grains in the simulated image begin to align with the respective bulk regions in the experiment, however, the positioning of the simulated interface relative to the interface in the experimental image is often unsuitable. In this case the simulated interface it is one row off. **(c)** For FOM values < 0.5, translative offset is increasingly rare, and the geometric fit appears better optimized. **(d)** For very low FOM values (< 0.2), the matching between the actual simulated and experimental image content is usually impressive (i.e. size of the atomic columns and the amount of blur match are better optimized).

With this, the goal of optimization is to find a set of parameters for the forward model, $\theta$, that produce a simulated image that can be arranged inside the experimental image in such a way that minimizes the objective, referred to as the figure-of-merit (FOM):

$$FOM(\theta) = \alpha\, d_{trans}(\theta) + \beta\, d_{sim}(\theta) \qquad (1)$$

where $d_{trans}(\theta)$ is the translation offset computed during fine alignment, $d_{sim}(\theta)$ is the similarity distance (the default is one minus the Structural Similarity Index Measure (SSIM) [41]), which quantifies the visual similarity between the simulation and experiment patch, and $\alpha$, $\beta$ are weights chosen to balance importance of each criteria ($\alpha = 0.1$, $\beta=1$ are default values). This FOM balances the importance of *geometric consistency* across the boundary of the simulated image, $d_{trans}$ (i.e. the atomic columns at the boundaries of the simulation are the same size, shape, and have the same orientation as the experimental columns they are replacing), and image *content consistency* within the boundaries of the simulated image $d_{sim}(\theta)$ (*i.e.* the shapes and relative intensities are aligned and self-consistent across images). With the default weights and the default SSIM similarity for $d_{sim}(\theta)$, we find the following approximate interpretation of the FOM values to hold for many of the samples observed: for FOM≥1, a significant translative offset with respect to the simulation/experiment boundary exists, usually due to scaling discrepancies or local distortions; for 1>FOM≥ 0.5, smaller translative offsets, if any, remain (typically, at least one of the bulk regions is well aligned with its respective experimental counterpart); for FOM<0.5, translative offset is increasingly rare, and the similarities between simulation and experiment image content is often noticeable. The solutions with FOM values < 0.2 represent the highest quality matches. In some cases, values this low are not attainable as much of this is influenced by the quality of the initial experimental input and level of structural disorder in the images sample. These observations are summarized in Figure 2.

We note that if an experimental image is exceptionally clear and the interface can be obtained by a simple geometric combination of the grains, the resulting structure potentially matches across all portions of the image (*i.e.,* both bulk regions and the interface). However, this is often not the case, and further local structure operations are needed to match the geometry of the interface more precisely. Moreover, a truly accurate representation of sample depth would involve further comparison with multislice simulations. Therefore, the process of obtaining these *ingrained* structures is considered an approximation or initial step, as opposed to exact structure determination.

*Output*. While the optimization proceeds, the default setting is for the current parameter set and FOM score to print to screen (and is also recorded in a *progress.out* file so that specific iterations can be revisited). An example of the optimization progress information provided is included in the following snippet:

```
Iteration 1:
        • pix_size (Å)             :  0.125
        • interface width (Å)      :  0.0
        • defocus (Å)              :  1.0
        • (x, y) shear (frac)      :  (0.0, 0.0)
        • (x, y) stretch (frac)    :  (0.0, 0.0)
        • img_size (pixels)        :  (285, 171)
        🌀 FOM                     :  0.6228565824012005
Iteration 2:
Warning — Solution contains significant translative offset (dtrans = 8)
        • pix_size (Å)             :  0.4
        • interface width (Å)      :  0.0
        • defocus (Å)              :  1.0
        • (x, y) shear (frac)      :  (0.0, 0.0)
        • (x, y) stretch (frac)    :  (0.0, 0.0)
        • img_size (pixels)        :  (129, 129)
        🌀 FOM                     :  2.186105511593367
```

At the end of the optimization, both the structure whose image was simulated, and the parameters used to fuse the images together are accessible. These parameters are valuable because they codify the transformation needed to go from an atomic structure to a simulated image, that now has an *explicit* association to the experimental image. In the next sections, we highlight applications involving structure and simulation/experiment correspondence output from *ingrained*. Further tools are included as part of the main repository that allow users to create videos from selected iterations recorded in *progress.out*

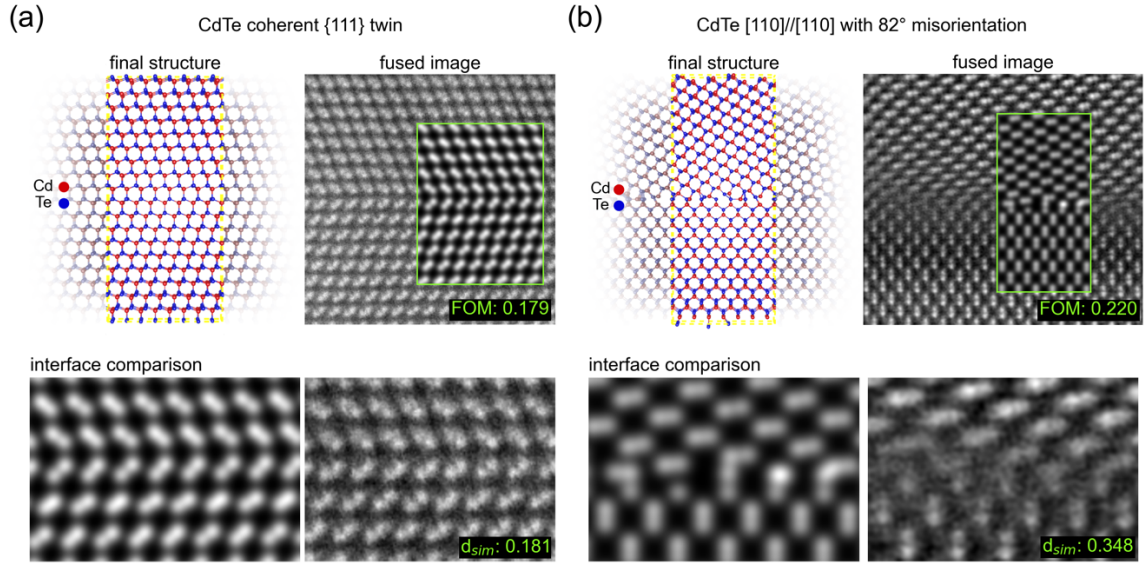

**Figure 3. | Coherent and incoherent grain boundaries in CdTe**. **(a)** The experimental image for the first collection of CdTe results contains a coherent {111} twin boundary viewed along the <110> direction. The final structure is given with the periodicity along the width highlighted. The fused image has a very low FOM score, indicating a high-quality fusion, which is confirmed in a comparison of the simulated and experimental interfaces. **(b)** The experimental image for the second collection of CdTe results contains an incoherent [110]//[110] tilt boundary with 82° misorientation angle. The quality of the resulting structure – as far as matching the bulk regions – is high, and even with the natural ambiguity of the interface structure, the simulation at the interface maintains close visual resemblance. The experimental images were obtained from the authors of [43]

**Methods**

The following examples showcase the capabilities of ingrained as both a tool for finding useful approximations of the structures in experimental imaging (for grain boundary and interfaces in particular), as well as for the development of experimentally-rationalized forward models in materials imaging. The case presented for forward model development involves STM simulation.

*Case #1: Coherent and incoherent grain boundaries in CdTe*

In this first example, we show one of the more straightforward structure initializations: a coherent {111} twin boundary in cadmium telluride (CdTe). The configuration file specifies a viewing direction along <110> and zinc-blende "F-43m" to form the bulk (to distinguish it from wurtzite and other less common metastable phases included in the MP database). In general, twin boundaries are extremely common in crystalline materials, and often form readily in response to thermal stress or applied deformation [42]. Figure 3a illustrates the resulting structure alongside both the final fused image and the interface comparison that results from the optimized fit-to-experiment. Because the crystals share a common plane of lattice points and mirror each other on either side of the interface, the resulting structure requires no strain to achieve coincidence along the width or depth and matches all portions of the image. For this reason, both the overall FOM and $d_{sim}$ values for the interfaces are particularly low (*e.g.,* < 0.20).

In the second example, we again use zinc-blende CdTe viewed along the <110> direction, but since the interface is now incoherent, the over lattice must be strained along the width to create a structure that is both computationally useful (in context of energetic calculations) but still fully periodic. Here, the magnitude of the strain is ~ 1%. Fortunately, since the viewing direction is common and the compound identical for both grains, like the twin boundary, this structure can also be constructed without strain along the depth. By observing diffraction patterns of the bulk crystalline regions, the misorientation angle between the crystals is measured at 82 degrees and is used to specify the tilt of the top in relation to the bottom. Figure 3b highlights a remarkable fit between simulation and experiment at the conclusion of image fusion, notwithstanding the unresolved structural details of the experimental interface. Guo *et al.* [43] use this initial structure to explore the role of Se and Cl segregation in the reduction of midgap states in CdSeTe, and even after DFT relaxation is used to further optimize the interface, the initial correspondence established by *ingrained* is still applicable.

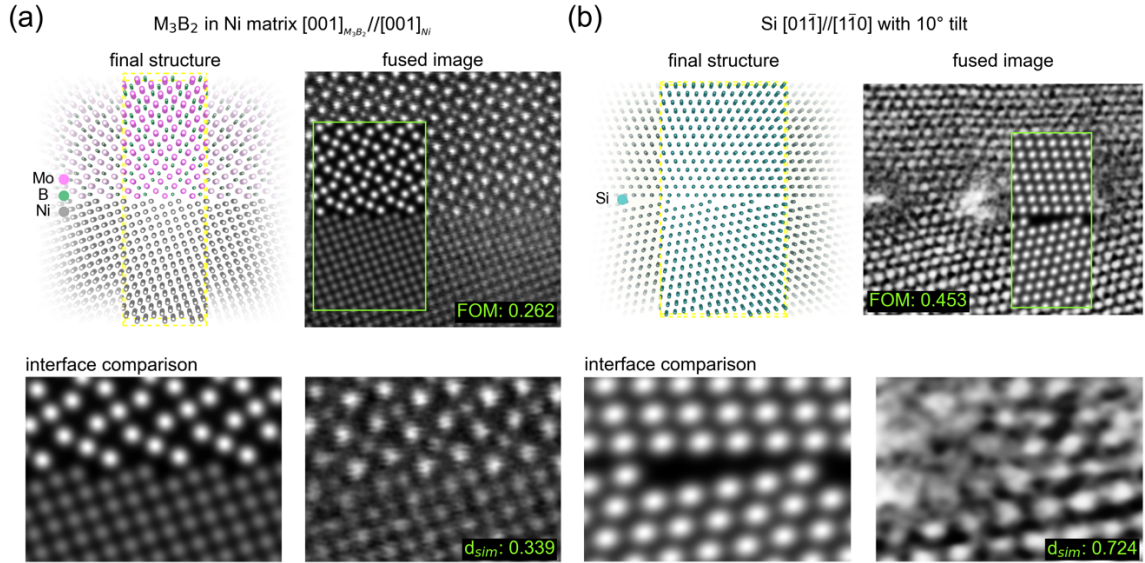

**Figure 4. | Interphase interfaces and grain boundaries with significant localized strain (a)** The experimental image associated with the image fusion contains an interface of an interphase $M_3B_2$ boride precipitate in a Ni-based super alloy ($[001]_{M3B2}$ //$[001]_{Ni}$). The overall FOM score reflects excellent geometric consistency across the boundary and within the image despite some inconsistencies in the relative intensities of the atomic columns. The experimental image was obtained from the authors of [44] **(b)** The image fusion based on a tilt grain boundary in Si [01-1]//[1-10] illustrates the difficulty in fitting a simulation with rigid affine transformations when the experimental image to be matched to contains significant local distortion and structural ambiguity at the interface.

*Case #2: Interphase interfaces and grain boundaries with significant localized strain*

In the previous CdTe examples, both crystalline grains were identical, and except for the presence of in-plane tilt in the incoherent case, this greatly simplified construction of the periodic over lattice. The collection of results in Figure 4a demonstrates how ingrained can be used to confirm the geometric compatibility of a specific boride precipitate/Ni matrix interface in an experimentally observed structure [44], where lattice mismatch must be resolved along both the width and depth dimensions. A $M_3B_2$-type precipitate/Ni interface (where "M" a transition metal element i.e. Cr, W, Mo, etc.) can be constructed from the tetragonal (P$_4$/mbm) Mo$_3$B$_2$ structure available in the Materials Project database. The specific structure tested in Figure 4a is the interface between (010) Mo$_3$B$_2$ and (1-30) Ni (viewed along <001> in each respective crystal). The ability to programmatically test geometric compatibility between different crystal phase in context of an experimentally observed structure is particularly useful for studies investigating complex interphase interfaces. Despite an excellent geometric fit (FOM = 0.262), one can observe that the light-heavy pattern of column intensities in the experimental image, is not well-reproduced in the simulation. This suggests that the Mo$_3$B$_2$ structure, though a good candidate from the perspective of geometric compatibility, is likely not consistent with the chemistry of the structure observed. It is possible that these intensity discrepancies would be resolved with higher accuracy multislice simulations, or perhaps this suggests that the sample has mixed cations in the M-site. Once a structure is suggested by ingrained, sampling of different cation orderings can be performed on the structure, which do not affect the overall alignment.

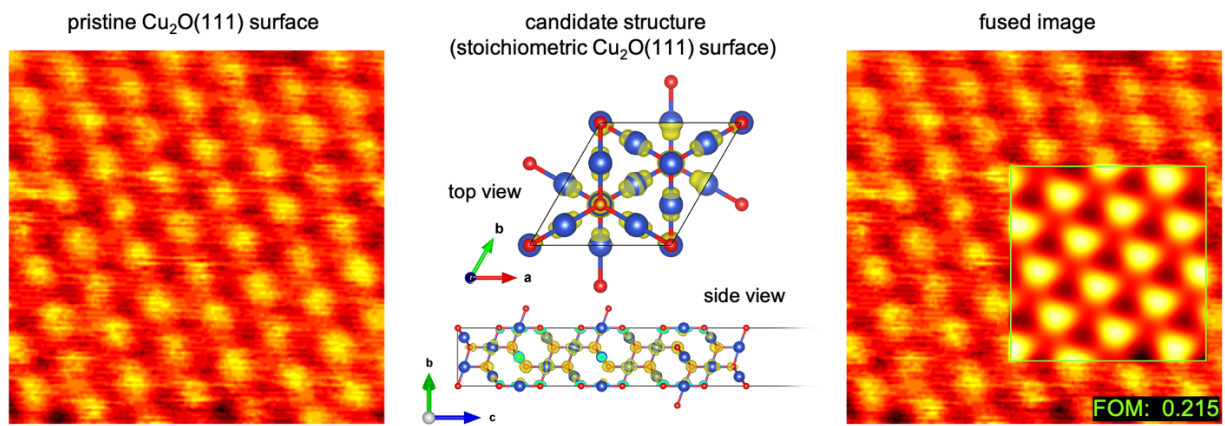

**Figure 5. | Confirmation of Cu₂O (111) surface with *ingrained* STM mode.** A high resolution STM image showing a pristine Cu$_2$O (111) surface is used as the experimental input to the *ingrained* framework. A DFT calculation for the proposed candidate structure, in the center, is used to create a simulated image of the surface, and *ingrained* confirms that the proposed structure is in fact consistent with experimental image, as described in [11].

The final collection of results, illustrated in Figure 4b, is based on a HAADF STEM image of a low-angle grain boundary in a thin Si nanowire [45]. Notice a significant amount of localized strain, blurring at the interface, and the reduced spatial resolution and signal levels (compared with some of the experimental images presented in the previous examples). Image registration, as implemented, only accounts for rigid affine transformation between the simulation and experiment. Therefore, a coarse association between simulation and experiment can still be made, but only to the extent that a rigid transformation applied to simulation can capture some these distinctly localized distortions. In the case of the Si nanowire in Figure 4b, the overall fit is perhaps adequate given the conditions, however, the similarity distance at the interface is unsurprisingly high ($d_{sim}$ = 0.724). Though further local structure manipulations and energetic calculations are necessary to better capture the non-rigid characteristics observed particularly around the interface, this does not diminish the value of having a decent approximate structure on which to base further analysis of the observed system.

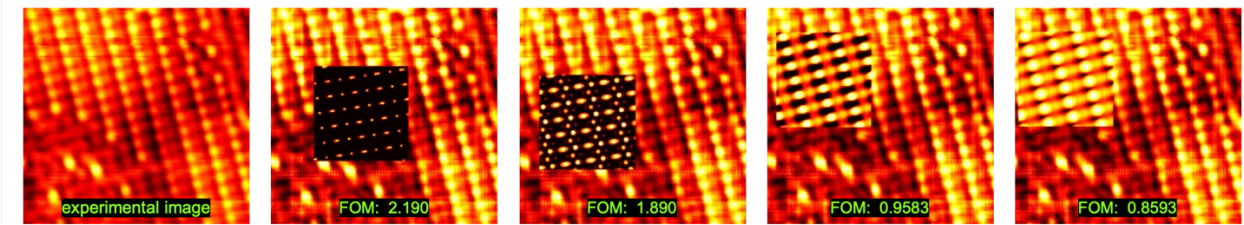

**Figure 6. | Optimization progress in search of borophene.** Ingrained applied on an STM image of rectangular hydrogenated borophene (borophane) [30]. The progression of snapshots taken in the course of optimization show improvement in both image structure (parameterization of the forward model) and in overall registration, suggesting that minimizing FOM values is sufficient for capturing fit-to-experiment. The optimized STM simulation from *ingrained* and excellent experimental match was taken as evidence in support of the proposed borophane structure [30]

*Case #3: STM mode and simulation parameter optimization for $Cu_2O$ (111)*

In the previous structure initialization cases, the registration of the HAADF STEM simulation to experiment was used to verify the geometry of a plausible bicrystal structure, much of which was decided by the selection of the specific grain chemistries, orientations, and tilt outside of the *ingrained* optimization. Updates to the forward modelling parameters had minimal effect on the geometric appearance of the imaged structure. This is because the observed intensity peaks were always consistent with the presence of an atomic column in the atomistic structure, and the image formation parameters only really served to adjust the height and width of those peaks. This was the assumption when convolution imaging was basis of the forward model. For other imaging modalities, the observed intensity is not always consistent with the presence of an explicit atomic column at that site, and what is visible instead complicates the interpretation of the intensities in the image. For example, artifacts such as halos, and shade-off, commonly observed in phase contrast microscopy, complicate edge interpretation in that the appearance of an edge in the image does not necessarily mean that a true object edge exists at that location [46]. In these instances, a simulated fit-to-experiment is necessary for rationalizing the image output of the forward model.

In this example, we examined STM images of a pristine $Cu_2O$(111) surface and proposed structure variants from *Zhang et al.* [11]. Using *ingrained* in combination with the DFT calculated partial charge densities, we were able to confirm that the atomic structure of the pristine (111) surface of a $Cu_2O$ bulk crystal was consistent with experimental image. The partial charge densities (PARCHG) near the Fermi level of a surface slab can be obtained through DFT calculations using widely used VASP code. Figure 5 shows the experimental image, proposed structure, and the final image fusion with the optimized imaging parameters. The parameterized fit between simulation and experiment is similar to what was outlined in *Zhang et al.* [11].

*Case #4: STM mode drives materials discovery*

The prior STM case provided validation of *ingrained* based on a known structure. When an initial structure is unknown, a multi-start approach, which implies that a variety of initial parameter configurations are tested on a population of candidate structures, can be used as means to filter out or focus in on certain structures of interest. In the case of STM, partial charge density information from several DFT calculations is the requisite "population" input, and the structures exhibiting the smallest

FOM values are interpreted as the most likely candidates. For example, Figure 6 depicts a progression of visual image improvements obtained through iteration of the *ingrained,* during the search for a hydrogenated borophene structure [30]. Among several candidate structures, the rectangular-2H model reported in the study, showed the lowest FOM which helped support its identification as the structure of rectangular borophane.

**Conclusion**

Formulating materials imaging simulations in such a way as to corroborate fundamental and nuanced aspects of experimental imaging is a critical challenge that must be addressed to fully harness the power of simulation and modeling in context of materials characterization. The *ingrained* framework presented here is a tool for atomic-resolution imaging that helps establish this simulated fit-to-experiment in an automated and robust way, using a coarse-to-fine image registration approach cast as an iterative optimization problem. Through examples of STEM images of grain boundaries and interfaces, and STM images of a surface, we showcase the power of *ingrained*, not only in its ability to forge an explicit association between simulation and experiment, but also in its versatility (*i.e.* numerous different imaging tasks can be improved with this approach). All the code for *ingrained* and the example cases explored in this work is available on GitHub. It is our hope that both computational researcher and microscopists alike will find practical use cases to add to the existing collection of examples outlined.

**Acknowledgments**

This work was supported, in part, as a part of the Center for Electrochemical Energy Science, an Energy Frontier Research Center funded by the US Department of Energy, Office of Science, Basic Energy Sciences under award number DE-AC02–06CH11. This work was performed, in part, at the Center for Nanoscale Materials, a U.S. Department of Energy Office of Science User Facility, and supported by the U.S. Department of Energy, Office of Science, under Contract No. DE-AC02-06CH11357. This work was also partially supported by the National Science Foundation Materials Research Science and Engineering Center at Northwestern University (NSF DMR-1720139). M.C. acknowledges the support from the BES SUFD Early Career award. This research used resources of the National Energy Research Scientific Computing Center, a DOE Office of Science User Facility supported by the Office of Science of the U.S. Department of Energy under Contract No. DE-AC02-05CH11231. We gratefully acknowledge the computing resources provided on Bebop, a high-performance computing cluster operated by the Laboratory Computing Resource Center at Argonne National Laboratory.

**References**

[1]   J. M. Cowley and A. F. Moodie, "The scattering of electrons by atoms and crystals. I. A new theoretical approach," *Acta Crystallogr.*, 1957.

[2]   A. Rosenauer *et al.*, "Measurement of specimen thickness and composition in Alx Ga1 - x N / GaN using high-angle annular dark field images," *Ultramicroscopy*, 2009.

[3]   T. Grieb *et al.*, "Determination of the chemical composition of GaNAs using STEM HAADF imaging and STEM strain state analysis," *Ultramicroscopy*, 2012.

[4]   A. De Backer, G. T. Martinez, A. Rosenauer, and S. Van Aert, "Atom counting in HAADF STEM

using a statistical model-based approach: Methodology, possibilities, and inherent limitations," *Ultramicroscopy*, 2013.

[5] G. T. Martinez, A. Rosenauer, A. De Backer, J. Verbeeck, and S. Van Aert, "Quantitative composition determination at the atomic level using model-based high-angle annular dark field scanning transmission electron microscopy," *Ultramicroscopy*, 2014.

[6] L. Duschek *et al.*, "Composition determination of semiconductor alloys towards atomic accuracy by HAADF-STEM," *Ultramicroscopy*, 2019.

[7] T. Paulauskas *et al.*, "Stabilization of a monolayer tellurene phase at CdTe interfaces," *Nanoscale*, 2019.

[8] M. Pedersen, M. L. Bocquet, P. Sautet, E. Lægsgaard, I. Stensgaard, and F. Besenbacher, "CO on Pt(111): Binding site assignment from the interplay between measured and calculated STM images," *Chem. Phys. Lett.*, 1999.

[9] F. Esch *et al.*, "Electron localization determines defect formation on ceria substrates," *Science (80-. )*, 2005.

[10] L. Li *et al.*, "Imaging Catalytic Activation of $CO_2$ on $Cu_2O$ (110): A First-Principles Study," *Chem. Mater.*, 2018.

[11] R. Zhang *et al.*, "Atomistic determination of the surface structure of $Cu_2O$(111): Experiment and theory," *Phys. Chem. Chem. Phys.*, 2018.

[12] M. Gong, S. Zhao, L. Jiao, D. Tian, and S. Wang, "A novel coarse-to-fine scheme for automatic image registration based on SIFT and mutual information," *IEEE Trans. Geosci. Remote Sens.*, 2014.

[13] D. L. G. Hill, P. G. Batchelor, M. Holden, and D. J. Hawkes, "Medical image registration," *Physics in Medicine and Biology*. 2001.

[14] K. S. Arun, T. S. Huang, and S. D. Blostein, "Least-Squares Fitting of Two 3-D Point Sets," *IEEE Trans. Pattern Anal. Mach. Intell.*, 1987.

[15] S. Brandt, J. Heikkonen, and P. Engelhardt, "Multiphase method for automatic alignment of transmission electron microscope images using markers," *J. Struct. Biol.*, 2001.

[16] S. Mohammadian *et al.*, "High accuracy, fiducial marker-based image registration of correlative microscopy images," *Sci. Rep.*, 2019.

[17] S. Somnath, C. R. Smith, N. Laanait, R. K. Vasudevan, and S. Jesse, "USID and Pycroscopy – Open Source Frameworks for Storing and Analyzing Imaging and Spectroscopy Data," *Microsc. Microanal.*, 2019.

[18] R. Bourne and R. Bourne, "ImageJ," in *Fundamentals of Digital Imaging in Medicine*, 2010.

[19] G. Lowe, "SIFT - The Scale Invariant Feature Transform," *Int. J.*, 2004.

[20] M. A. Fischler and R. C. Bolles, "Random sample consensus: A Paradigm for Model Fitting with Applications to Image Analysis and Automated Cartography," *Commun. ACM*, 1981.

[21] M. Guizar-Sicairos, S. T. Thurman, and J. R. Fienup, "Efficient subpixel image registration algorithms," *Opt. Lett.*, 2008.

[22] Y. Douini, J. Riffi, A. M. Mahraz, and H. Tairi, "An image registration algorithm based on phase correlation and the classical Lucas–Kanade technique," *Signal, Image Video Process.*, 2017.

[23] L. Jones *et al.*, "Smart Align—a new tool for robust non-rigid registration of scanning microscope data," *Adv. Struct. Chem. Imaging*, 2015.

[24] J. M. Fitzpatrick, D. L. G. Hill, and C. R. Maurer Jr., "Chapter 8: Image Registration," *Handb. Med. Imaging Vol 2*, 2000.

[25] P. Thévenaz, U. E. Ruttimann, and M. Unser, "A pyramid approach to subpixel registration based on intensity," *IEEE Trans. Image Process.*, 1998.

[26] A. B. Yankovich *et al.*, "Picometre-precision analysis of scanning transmission electron microscopy images of platinum nanocatalysts," *Nat. Commun.*, 2014.

[27] Y. Wang *et al.*, "Correcting the linear and nonlinear distortions for atomically resolved STEM spectrum and diffraction imaging," *Microscopy*, 2018.

[28] O. Clatz *et al.*, "Robust nonrigid registration to capture brain shift from intraoperative MRI," *IEEE Trans. Med. Imaging*, 2005.

[29] J. Tersoff and D. R. Hamann, "Theory of the scanning tunneling microscope," *Phys. Rev. B*, 1985.

[30] Q. Li *et al.*, "Synthesis of borophane polymorphs through hydrogenation of borophene," *Science*, vol. 371, no. 6534, pp. 1143–1148, Mar. 2021.

[31] A. Jain *et al.*, "The Materials Project: A materials genome approach to accelerating materials innovation," *APL Mater.*, vol. 1, no. 1, p. 11002, 2013.

[32] C. Buurma *et al.*, "Creation and analysis of atomic structures for CdTe bi-crystal interfaces by the grain boundary genie," in *2015 IEEE 42nd Photovoltaic Specialist Conference, PVSC 2015*, 2015.

[33] E. J. Kirkland, *Advanced computing in electron microscopy: Second edition*. 2010.

[34] C. Ophus, "A fast image simulation algorithm for scanning transmission electron microscopy," *Adv. Struct. Chem. imaging*, vol. 3, no. 1, p. 13, 2017.

[35] A. Pryor, C. Ophus, and J. Miao, "A streaming multi-GPU implementation of image simulation algorithms for scanning transmission electron microscopy," *Adv. Struct. Chem. imaging*, vol. 3, no. 1, p. 15, 2017.

[36] G. Kresse and J. Furthmüller, "Efficiency of ab-initio total energy calculations for metals and semiconductors using a plane-wave basis set," *Comput. Mater. Sci.*, 1996.

[37] G. Kresse and J. Furthmüller, "Efficient iterative schemes for ab initio total-energy calculations using a plane-wave basis set," *Phys. Rev. B - Condens. Matter Mater. Phys.*, 1996.

[38] S. Van Der Walt *et al.*, "Scikit-image: Image processing in python," *PeerJ*, 2014.

[39] M. J. D. Powell, "An efficient method for finding the minimum of a function of several variables without calculating derivatives," *Comput. J.*, 1964.

[40] P. Virtanen *et al.*, "SciPy 1.0: fundamental algorithms for scientific computing in Python," *Nat. Methods*, 2020.

[41] Z. Wang, A. C. Bovik, H. R. Sheikh, E. P. Simoncelli, and others, "Image quality assessment: from error visibility to structural similarity," *IEEE Trans. image Process.*, vol. 13, no. 4, pp. 600–612, 2004.

[42] A. Kelly and K. M. Knowles, *Crystallography and Crystal Defects: Second Edition*. 2012.

[43] J. Guo *et al.*, "Effect of selenium and chlorine co-passivation in polycrystalline CdSeTe devices," *Appl. Phys. Lett.*, 2019.

[44] X. B. Hu *et al.*, "Atomic-Scale Investigation of the Borides Precipitated in a Transient Liquid Phase-Bonded Ni-Based Superalloy," *Metall. Mater. Trans. A Phys. Metall. Mater. Sci.*, 2020.

[45] Z. Sun *et al.*, "Strain-Energy Release in Bent Semiconductor Nanowires Occurring by Polygonization or Nanocrack Formation," *ACS Nano*, vol. 13, no. 3, pp. 3730–3738, Mar. 2019.

[46] Z. Yin, T. Kanade, and M. Chen, "Understanding the phase contrast optics to restore artifact-free microscopy images for segmentation," *Med. Image Anal.*, 2012.